\documentclass[twocolumn,english,aps,prl,superscriptaddress]{revtex4-2}
\usepackage[T1]{fontenc}
\usepackage[utf8]{inputenc}
\usepackage{amsmath}
\usepackage{amssymb}
\usepackage{graphicx}
\usepackage{esint}

\makeatletter
\usepackage{xcolor}
\usepackage{url}
\usepackage[colorlinks,citecolor=blue,linkcolor=blue]{hyperref}

\makeatother

\usepackage{babel}
\begin{document}
\title{Power-Efficiency Constraint for Chemical Motors}
\author{R. X. Zhai}
\address{Graduate School of China Academy of Engineering Physics, Beijing,
100193, China}
\author{Hui Dong}
\email{hdong@gscaep.ac.cn}

\address{Graduate School of China Academy of Engineering Physics, Beijing,
100193, China}
\date{\today}
\begin{abstract}
Chemical gradients provide the primordial energy for biological functions
by driving the mechanical movement of microscopic engines. Their thermodynamic
properties remain elusive, especially concerning the dynamic change
in energy demand in biological systems. In this article, we derive
a constraint relation between the output power and the conversion
efficiency for a chemically fueled steady-state rotary motor analogous
to the $\mathrm{F}_{0}$-motor of ATPase. We find that the efficiency
at maximum power is half of the maximum quasi static efficiency. These
findings shall aid in the understanding of natural chemical engines
and inspire the manual design and control of chemically fueled microscale
engines.
\end{abstract}
\maketitle

\section{Introduction}

As a counterpart to heat engines \citep{Huang1987}, chemical engines
are fueled by chemical reactions, converting chemical energy into
mechanical work \citep{chen_performance_1997,Schmiedl2007,sieniutycz_analysis_2008,Amano2021,hooyberghs_efficiency_2013,leighton_inferring_2023,Leighton2024}.
These engines, such as the Kinesin and ATP synthase, are ubiquitous
in open systems like living organisms and play crucial roles \citep{dimroth_energy_1999,hackney_enzymes_2004,von_ballmoos_essentials_2009,phillips_physical_2013}.
Recently, the thermodynamic behavior of these engines, together with
the related chemical processes, has garnered significant attention
\citep{Rao2016,wilson_autonomous_2016,Amano2021,Amano2022,lathouwers_internal_2022,Borsley2022,Singh2023,Leighton2024},
especially on the power-efficiency trade-off of chemical engines in
finite time due to the substantial varying energy consumption rate
in biological systems.

Efficiency, which measures the conversion ratio of free energy to
mechanical work, is a core parameter of thermodynamic engines. However,
the optimal efficiency is typically achieved only when the cycle time
approaches infinity resulting in a zero power output \citep{curzon_efficiency_1975,VandenBroeck2005,Tu2008JPhysAMathTheor41_312003,esposito_efficiency_2010,Lin2021},
which is another critical parameter for assessing the performance
of chemical engines. The pursuit of non-zero power inevitably leads
to a decrease in efficiency, making the trade off \citep{Holubec2016,Holubec2017,Holubec2018,Proesmans2016,TradeoffrelationShiraishi,Pietzonka2018,Constraintrelationyhma,zhai_experimental_2023}
between power and efficiency a significant topic in finite-time thermodynamics
\citep{Andresen1984,seifert_stochastic_2012}. For chemical engines,
quantitatively studying the power-efficiency constraint will establish
performance boundaries, aiding in the understanding of natural chemical
engines \citep{julicher_modeling_1997,bustamante_physics_2001,Wang2002,Golubeva2012,Schmiedl2008,Bauer2008,zhang_efficiency_2017,leighton_inferring_2023}
and inspiring the manual design and control of microscale engines
fueled by chemical reactions \citep{wilson_autonomous_2016,Raz2016,Borsley2022,Amano2022,Leighton2024}.

In this article, we investigate a rotary motor fueled by chemical
reactions, resembling the $\mathrm{F}_{o}$ motor of ATP synthase
\citep{dimroth_energy_1999,hackney_enzymes_2004,von_ballmoos_essentials_2009}.
We use thermodynamic cycles to describe the operation of such steady-state
engine, which interacts with two reservoirs with different chemical
potentials, performing mechanical work through its autonomous rotation.
Its power and efficiency are regulated by its rotation speed and an
inside potential. Our analysis reveals the power-efficiency constraint
relation for this chemical engine. When the chemical potential difference
between the two reservoirs is small, this constraint takes a simple
analytical form. Interestingly, in this regime, the ratio between
efficiency at maximum power and the upper limit of efficiency is $1/2$,
a coefficient that appears to be universal for thermodynamic cycles
in the linear response regime.

\begin{figure}
\begin{centering}
\includegraphics[width=7.3cm]{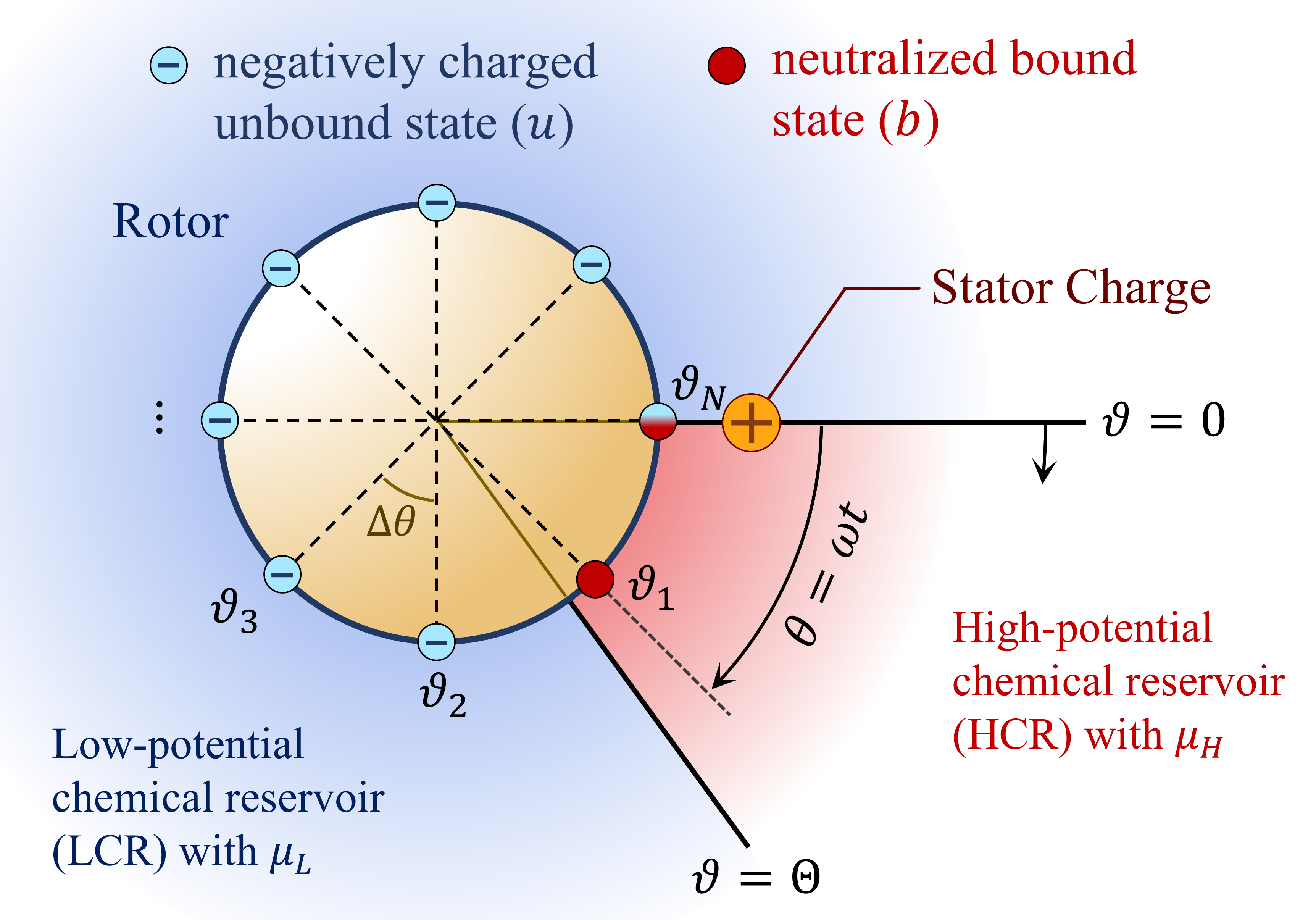}
\par\end{centering}
\caption{\protect\label{fig:ATPase_model} The rotary motor model consisting
of a rotor and a stator. The rotor has a ring structure composed of
multiple c-subunits, depicted by yellow blocks. Each subunit contains
a negatively charged ionic binding site with the light blue representing
the unbound state ($u$) with negative charge and dark red as the
neutralized bound state $(b)$. The rotation coordinate of the rotor
is denoted as $\theta$, where a clockwise rotation is defined as
the positive direction. The stator interacts with the rotor through
a fixed charge, represented by the large orange circle, located at
$\theta=0$. This charge provides an attractive inside potential $\Phi(\theta)$
for the negatively charged sites. The motor interacts with a high
chemical potential ($\mu_{H}$) reservoir (HCR) (red shaded region)
with the angular width $\Theta$, and a low chemical potential ($\mu_{L}$)
reservoir (LCR) (blue shaded region).}
\end{figure}

\section{Model of biological rotary motor}

Fig. \ref{fig:ATPase_model} illustrates the current rotary motor,
consisting of a rotor and a stator. The rotor comprises $N$ negatively
charged binding sites, which are evenly distributed along a circular
ring. The charge of these sites can be neutralized by binding with
one positively charged ion from the reservoirs, as described in the
following chemical reaction: 
\begin{equation}
\text{unbound state}+\mathrm{H^{+}}\rightleftharpoons\text{bound state}.\label{eq:reaction}
\end{equation}
The light blue and dark red colors on these sites correspond to the
neutralized bound state (denoted by $b$) and negatively charged unbound
state (denoted by $u$) respectively.

For the $j$-th binding site, its position $\vartheta_{j}$ is obtained
by $\vartheta_{j}=\theta+(j-1)\Delta\theta$, where $\Delta\theta=2\pi/N$.
The parameter $\theta$ represents the position of the first site
and can also serve as the generalized coordinate of the rotor's overall
rotation state. The state for all binding sites $(s_{1},\ldots,s_{N})=b,u$,
together with the coordinate $\theta$, constitute the configuration
space for the rotor $(\{s_{j}\},\theta)$.

When the motor achieve its steady state, the force generated by the
rotor at steady rotation is approximately constant due to the dense
and evenly-spaced binding sites, and is balanced by a constant load
and the friction, inducing a constant rotation speed $\dot{\theta}=\omega$.
Under this approximation, all binding sites exhibit identical temporal
evolution, aside from a phase delay. Consequently, studying the thermodynamic
cycle of one specific binding site, e.g., the first site, is sufficient
to reflect the behavior of the whole motor. For example, the work
for the whole motor during a complete cycle is obtained by multiplying
the work from one site with the number $N$. Our subsequent analysis
will focus on the thermodynamic cycle of the first binding site, whose
configuration space is expressed as $(s,\theta)$.

For the site of interest, its energy $U(u,\theta)$ in the unbound
state $u$ is determined by an inside potential $\Phi(\theta)$ due
to the Coulomb interaction between its charge and a fixed charge (the
large orange circle) on the stator located at $\vartheta_{0}=0$,
namely $U(u,\theta)=\Phi(\theta)$. We apply the Coulomb interaction
to determine the form of $\Phi(\theta)$, which is presented in Appendix
\ref{sec:Form-of-the}. The energy $U(b,\theta)$ for the site in
the bound state $b$ is given only by the binding energy $E_{b}$
without the Coulomb interaction, namely $U(b,\theta)=E_{b}$.

The motor is in contact with two chemical reservoirs: a high chemical
potential reservoir (HCR) with chemical potential $\mu_{H}$ (red
shadowed region around the motor) and a low chemical potential reservoir
(LCR) with chemical potential $\mu_{L}$ (blue shadowed region around
the motor). Thus, the environmental chemical potential for the binding
site of interest $\mu(\theta)$ is expressed as a function of its
position $\theta$, i.e., $\mu(\theta)=\mu_{H}$ for $0\le\theta<\Theta$,
and $\mu(\theta)=\mu_{L}$ for $\Theta<\theta\le2\pi$. Here, $\Theta$
is the angular width of the HCR. During the rotation, the binding
sites alternate their contact between the high and low potential reservoirs.

We denote for the site of interest the average binding number in the
bound state $b$ as $n$ ($0\le n\le1$). When the site is in equilibrium
with a reservoir, the average binding number $n=n^{(0)}$ is determined
by the energy decrease $\Delta U(\theta)=E_{b}-\Phi(\theta)-\mu(\theta)$
of reaction (\ref{eq:reaction}), as
\begin{equation}
n^{(0)}(\theta)=\frac{1}{1+\mathrm{e}^{\beta\Delta U(\theta)}},\label{eq:average_binding=000023_qs}
\end{equation}
where $\beta=(k_{B}T)^{-1}$ is the inverse temperature of the reservoirs.
For finite-time processes, the evolution of $n$ is determined by
the rate equation \citep{Rao2016} of reaction (\ref{eq:reaction}),
and the change rate of the average binding number $n$ is proportional
to its difference from the equilibrium, i.e., $\dot{n}=-\left(n-n^{(0)}\right)/\tau_{r}$,
where $\tau_{r}$ is the timescale of relaxation between the binding
site and the reservoir. Here, we have assumed the same timescale $\tau_{r}$
for the relaxation in the two chemical reservoirs. Replacing the variable
$t$ with the angular coordinate $\theta=\theta_{0}+\omega t$, we
obtain the evolution equation of $n(\theta)$ as
\begin{align}
\frac{\mathrm{d}n}{\mathrm{d}\theta} & =-\frac{1}{\omega\tau_{r}}\left(n-n^{(0)}(\text{\ensuremath{\theta}})\right),\label{eq:evolution_eq}
\end{align}
with a periodic boundary condition $n(\theta)=n(\theta+2\pi)$, the
effect of finite operation time is characterized by the dimensionless
coefficient $\omega\tau_{r}$.

\section{Power and efficiency of the motor}

The average energy of one site is expressed with the average binding
number as $F=nE_{b}+(1-n)\Phi(\theta)$ with its change given by $\mathrm{d}F=[E_{b}-\Phi(\theta)]\mathrm{d}n+(1-n)\Phi'(\theta)\mathrm{d}\theta.$
The first term is the energy change caused by the particle exchange
with the reservoirs, and the second term represents the mechanical
work $W_{\mathrm{mech}}=\int_{-\pi}^{\pi}(1-n)\Phi'(\theta)\mathrm{d}\theta$.
For rotation with a finite rotation speed $\omega$, we introduce
a linear dissipation $W_{\mathrm{diss}}=2\pi\gamma\omega$ to account
for the energy loss due to friction. Here $\gamma$ is the given linear
friction coefficient accounting for the resistance torque proportional
to the rotation speed $\omega$. The net work output per cycle is
obtained as $W_{\mathrm{net}}=W_{\mathrm{mech}}-W_{\mathrm{diss}}$.
The thermodynamic efficiency of the cycle is, 
\begin{equation}
\eta=\frac{W_{\mathrm{net}}}{\Delta\mu\Delta n},\label{eq:def_efficiency}
\end{equation}
where $\Delta\mu=\mu_{H}-\mu_{L}$ is the chemical potential difference,
and $\Delta n=n(\Theta)-n(0)$ is the average number of particles
transferred from the HCR to the LCR. Here, $\Delta\mu\Delta n$ describes
the free energy decrease for the chemical reservoirs due to the transport
of $\Delta n$ ions. The efficiency in Eq. (\ref{eq:def_efficiency})
reflects the ratio of the net work output $W_{\mathrm{net}}$ with
respect to the free energy decrease. The (average) power is obtained
as the average mechanical energy output rate per cycle
\begin{equation}
P=\frac{W_{\mathrm{net}}}{\tau_{\mathrm{cycle}}},\label{eq:power_definition}
\end{equation}
where $\tau_{\mathrm{cycle}}=2\pi/\omega$ is the cycle time.

\section{Maximum Quasi-static Efficiency}

\begin{figure*}[t]
\centering{}\includegraphics{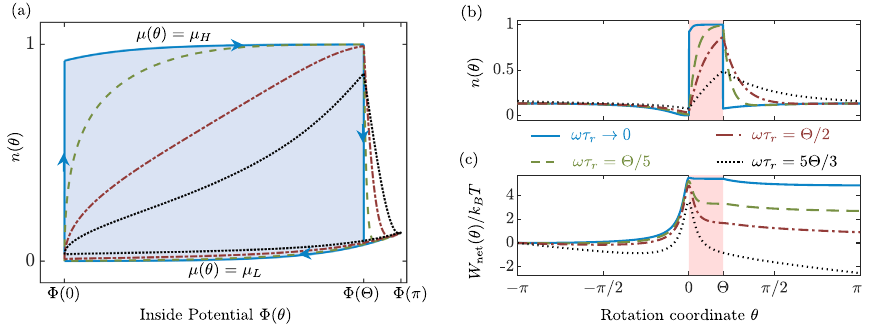}\caption{\protect\label{fig:fixedOmegaevolution} The thermodynamic cycle of
our biological rotary motor. (a) $n-\Phi$ diagram for the motor.
The area of the blue shaded region reflects the mechanical work output
for the quasi static cycle. (b-c) Average binding number $n$ and
the cumulative net work output $W_{\mathrm{net}}$ as functions of
$\theta$ within finite operation time. The red shaded region $(0\protect\leq\theta<\Theta)$
represents the range where the site is in contact with the high potential
chemical reservoir. The horizontal coordinates are the same for the
two subfigures. The parameters are fixed as follows: $\Theta=\pi/5$,
$V=5k_{B}T$, $\Delta\mu=10k_{B}T$ , and $\gamma=0.5\tau_{r}k_{B}T$.
The blue solid curve shows the evolution in the quasi static limit
with $\omega\rightarrow0$. The green dashed curves, red dash-dotted
curves and black dotted curves show the evolution with $\omega\tau_{r}=\Theta/5,\Theta/2,5\Theta/3$,
respectively. The increase in rotation speed causes a decrease in
total work output per cycle, which is reflected by the final value
of the curves $W_{\mathrm{net}}(\pi)$. It is noteworthy that for
$\omega\tau_{r}=5\Theta/3$, the total work output $W_{\mathrm{net}}$
becomes negative, which means the motor no longer works as an engine.}
\end{figure*}

In the quasi static limit with $\omega\tau_{r}\rightarrow0$, the
binding site keeps in equilibrium with the reservoir, and hence the
average binding number is obtained by Eq. (\ref{eq:average_binding=000023_qs}).
The energy dissipated due to the friction $W_{\mathrm{diss}}$ is
$0$ in such quasi static limit. The net work output of a complete
cycle for the quasi static limit is then obtained as 
\begin{align*}
W_{\mathrm{net}}^{(0)} & =-\int_{0}^{2\pi}\left(1-n(\theta)\right)\Phi'(\theta)\mathrm{d}\theta\\
 & =-\oint\left(1-n\right)\mathrm{d}\Phi=\oint n\mathrm{d}\Phi.
\end{align*}
In the $n-\Phi$ diagram shown in Fig. \ref{fig:fixedOmegaevolution}(a),
the net work output is the area of the light blue shaded region enclosed
by the blue solid line shown with $n(\theta),\Phi(\theta)$. The work
output is obtained as
\begin{align}
W_{\mathrm{net}}^{(0)}= & k_{B}T\ln\frac{1+\mathrm{e}^{\beta\left(\Phi(0)-E_{b}+\mu_{L}\right)}}{1+\mathrm{e}^{\beta\left(\Phi(0)-E_{b}+\mu_{H}\right)}}\nonumber \\
 & -k_{B}T\ln\frac{1+\mathrm{e}^{\beta\left(\Phi(\Theta)-E_{b}+\mu_{L}\right)}}{1+\mathrm{e}^{\beta\left(\Phi(\Theta)-E_{b}+\mu_{H}\right)}}.\label{eq:quasi-static-work}
\end{align}
We conclude that the quasi static work is determined by the values
of $\Phi(\theta)$ at the edges of the channel, i.e., $\Phi(0)$ and
$\Phi(\Theta)$. The number of particles transferred in a quasi static
cycle is given by 
\begin{equation}
\Delta n^{(0)}=\frac{1}{1+\mathrm{e}^{-\beta(\Phi(\Theta)-E_{b}+\mu_{H})}}-\frac{1}{1+\mathrm{e}^{-\beta(\Phi(0)-E_{b}+\mu_{L})}}.\label{eq:quasi-static-particlenumber}
\end{equation}
We first optimize $E_{b}$ to get the maximum quasi static efficiency
$\eta^{(0)}$ as derived from Eq. (\ref{eq:def_efficiency}), by requiring
that $\partial\eta^{(0)}/\partial E_{b}=0$. The optimum $E_{b}$
is obtained as
\[
E_{b}=\frac{\mu_{H}+\mu_{L}+\Phi(\Theta)+\Phi(0)}{2},
\]
with which the quasi static efficiency is maximized as

\begin{equation}
\eta_{\mathrm{max}}^{(0)}=\frac{1+\cosh(\beta\Delta_{+}/2)}{\beta\Delta\mu\sinh(\beta\Delta_{+}/2)}\ln\left(\frac{1+\cosh(\beta\Delta_{+}/2)}{1+\cosh(\beta\Delta_{-}/2)}\right),\label{eq:quasi-static-efficiency}
\end{equation}
where $\Delta_{+}\equiv V+\Delta\mu$, $\Delta_{-}\equiv V-\Delta\mu$,
and $V\equiv\Phi(\Theta)-\Phi(0)$ defines the inside potential depth.
In the quasi static limit, the detailed shape of the internal potential
$\Phi(\theta)$ is irrelevant, and only inside potential depth (IPD)
$V$ determines the overall efficiency of the cycle. The upper limit
of quasi static efficiency in Eq. (\ref{eq:quasi-static-efficiency})
is obtained as $1$ with $V\rightarrow\infty$ or $\Delta\mu\rightarrow0$,
representing that all the free energy drawn from the HCR is converted
into mechanical work. For further discussions about these limits,
please see Appendix \ref{sec:n_Phi_diag}. Such quasi static cycles
are impractical for biological systems, since the power output is
$0$ noting that the cycle time $\tau_{\mathrm{cycle}}$ is infinite.

\section{Constraint between power and efficiency}

To acquire a finite output power, the motor rotates with a non-zero
rotation speed $\omega$. The finite time behavior of the motor is
numerically evaluated in Fig. \ref{fig:fixedOmegaevolution}. The
blue solid curves illustrate the quasi static cycle, while the green
dashed curves, the red dash-dotted curves and the black dotted curves
illustrate cycles with rotation speeds $\omega\tau_{r}=\Theta/5$,
$\Theta/2$ and $5\Theta/3$, respectively. In Fig. \ref{fig:fixedOmegaevolution}
(a), the $n-\Phi$ diagrams are plotted. It is evident that the mechanical
work decreases with an increase in $\omega$. We note that for large
$\omega$ , e.g., $\omega\tau_{r}=5\Theta/3$, the total work output
$W_{\mathrm{net}}$ becomes negative, which means the motor no longer
works as an engine. In the following discussion, we only consider
the situation with positive output work $W_{\mathrm{net}}>0$.

The effect of such finite time operation will induce not only smaller
mechanical work, but also larger frictional dissipation $W_{\mathrm{diss}}$.
In Fig. \ref{fig:fixedOmegaevolution} (b) and (c), the curves show
the average binding number $n(\theta)$ (in Fig. \ref{fig:fixedOmegaevolution}(b))
and the cumulative net work output $W_{\mathrm{net}}(\theta)=\int_{-\pi}^{\theta}(\mathrm{d}W_{\mathrm{mech}}-\mathrm{d}W_{\mathrm{diss}})$
(in Fig. \ref{fig:fixedOmegaevolution}(c)) as functions of the rotational
coordinates $\theta$ for the range of $-\pi<\theta\le\pi$. The red-shaded
region in Figs. (b-c) illustrates the range where the binding site
is in contact with the HCR. With increasing rotation speed $\omega$,
the relaxation processes in the two chemical potential reservoirs
become inadequate. And in turn, this results in the decrease in the
total net output work. In the calculation, we set the width $\Theta=\pi/5$
and the dissipation coefficient $\gamma=0.5k_{B}T\tau_{r}$.

With the decrease of $\Delta\mu$, the mechanical work becomes smaller.
Thus, to maintain a positive work output, $\omega$ should also decrease
to reduce the dissipation. In such linear response (LR) regime with
both $\Delta\mu$ and $\omega\tau_{r}$ are infinitesimal, the net
work output and the exchanged particle number are expressed with series
in $\omega$ and $\Delta\mu$. Up to the second order of $\Delta\mu$,
the upper bound of finite time power output is obtained approximately
as
\begin{equation}
P\le4P_{\max}\eta(1-\eta).\label{eq:power_efficiency_constraint}
\end{equation}
 where the maximum power (MP) $P_{\mathrm{max}}$ reads

\begin{equation}
P_{\mathrm{max}}=\frac{\Delta\mu^{2}}{8\pi}\max\left(\frac{\left[\Omega^{(0)}(V)\right]^{2}}{2\pi\gamma+\Lambda^{(0)}(V)\tau_{r}/\beta}\right).\label{eq:MP}
\end{equation}
Here $\Omega^{(0)}(V)$ and $\Lambda^{(0)}(V)$ are functions of the
IPD $V$. The detailed derivation for this solution is shown in Appendix
\ref{sec:Power-efficiency-constraint}. This relation characterizes
the universal constraint between the power and efficiency of our motor
within the LR regime. The corresponding efficiency, typically known
as the efficiency at maximum power (EMP) $\eta_{\mathrm{MP}}=1/2$
is half of the maximum efficiency, being independent of the detailed
parameters and configuration of the motor. The coefficient $1/2$
appears to be universal in the LR regime for all thermodynamic cycles
\citep{VandenBroeck2005,esposito_universality_2009,esposito_efficiency_2010,Tu2012,hooyberghs_efficiency_2013,Josefsson2018,chenquan2022arxiv}.

\begin{figure}
\centering{}\includegraphics{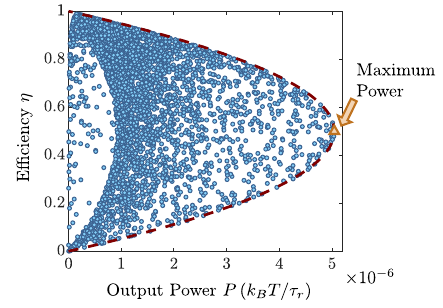}\caption{\protect\label{fig:Constraint_relation} The constraint between the
output power $P$ and the efficiency for $\Delta\mu=0.1k_{B}T$. We
randomly evaluate $10^{4}$ cycles with different combinations of
rotation speed and IPD $(\omega,V)$, whose power and efficiency are
shown with blue circles. The constraint relation within the LR regime
is illustrated by the region enclosed by the dark dashed curve. The
MP point with $P_{\mathrm{max}}=5.04\times10^{-6}k_{B}T/\tau_{r}$
and $\eta_{\mathrm{MP}}=0.50$ is indicated by the orange triangle.}
\end{figure}

To validate our constraint relation in Eq. (\ref{eq:power_efficiency_constraint}),
we evaluate the efficiency and power by numerically solving Eq. (\ref{eq:evolution_eq})
for different combinations of rotation speed and IPD $(\omega,V)$
with parameters $\Theta=\pi/5$, $\gamma=5k_{B}T\tau_{r}$, and $\Delta\mu=0.1k_{B}T$.
The results of $10^{4}$ random combinations with $0<V/k_{B}T<60$
and $0<\omega<W^{(0)}/2\pi\gamma$ are illustrated as blue circles
in Fig. \ref{fig:Constraint_relation}. The constraint relation in
Eq. (\ref{eq:power_efficiency_constraint}) is shown with the region
enclosed by the dark red dashed curve. Within the LR regime, the data
align well with the prediction, yielding the MP $P_{\mathrm{max}}=5.04\times10^{-6}k_{B}T/\tau_{r}$
and EMP $\eta_{\mathrm{MP}}=0.50$, indicated by the orange triangle.

\begin{figure}
\begin{centering}
\includegraphics{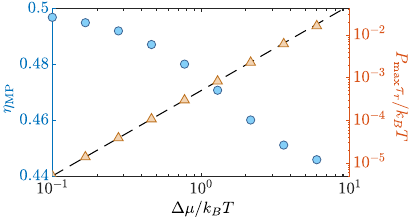}
\par\end{centering}
\caption{\protect\label{fig:Delta_mu_EMP} Efficiency at maximum power and
maximum power within the linear response regime. As functions of the
chemical gradient $\Delta\mu$, the efficiency at maximum power is
illustrated with blue circles, and the maximum power is illustrated
with yellow triangles. The dashed line shows the leading order proportional
to $\Delta\mu^{2}$ of the maximum power.}
\end{figure}

To quantitatively evaluate the behavior outside the LR regime, we
present the EMP (blue circles) and the MP (orange triangles) as functions
of $\Delta\mu$ in Fig. \ref{fig:Delta_mu_EMP} for $\Theta=\pi/5$
and $\gamma=5k_{B}T\tau_{r}$. The dashed line shows the maximum power
obtained from Eq. (\ref{eq:MP}). The efficiency at maximum power
$\eta_{\mathrm{MP}}$ decreases with increasing chemical gradient
$\Delta\mu$, while the power increases with the increasing chemical
gradients $\Delta\mu$. We note that the theoretical prediction in
Eq. (\ref{eq:MP}) for the LR regime also fits well with the simulation
beyond the LR regime.

\section{Conclusion}

We would like to mention that much attention has been drawn to the
thermodynamic properties of the chemically fueled rotary engines,
with most studies concentrating on the $F_{1}$ motor and the interaction
between the $F_{1}$ and $F_{0}$ parts. Yet, the thermodynamic cycle
for the $F_{0}$ portion is seldom explored, especially for its dynamic
impact on power and efficiency. In this article, we have evaluated
the constraint between the power and efficiency of a chemically fueled
rotary motor operating at a steady rotation, considering various parameters
such as the rotation speed, the depth of the inside potential and
the friction coefficient. With the constraint relation, the maximum
power and corresponding efficiency at maximum power are acquired.
The efficiency at maximum power approaches $1/2$ in the linear response
regime, which appears to be a universal coefficient for all finite-time
thermodynamic cycles. We anticipate that this research will inspire
further investigations into the thermodynamics of microscopic motors.

\section*{Acknowledgments}

R. X. Zhai is grateful to Y. H. Ma for his discussions. This work
is supported by the Innovation Program for Quantum Science and Technology
(Grant No. 2023ZD0300700) and the National Natural Science Foundation
of China (Grants No. U2230203, No. U2330401, and No. 12088101).

\appendix

\section{Form of the inside Potential}

\begin{figure}[!h]
\centering{}\includegraphics{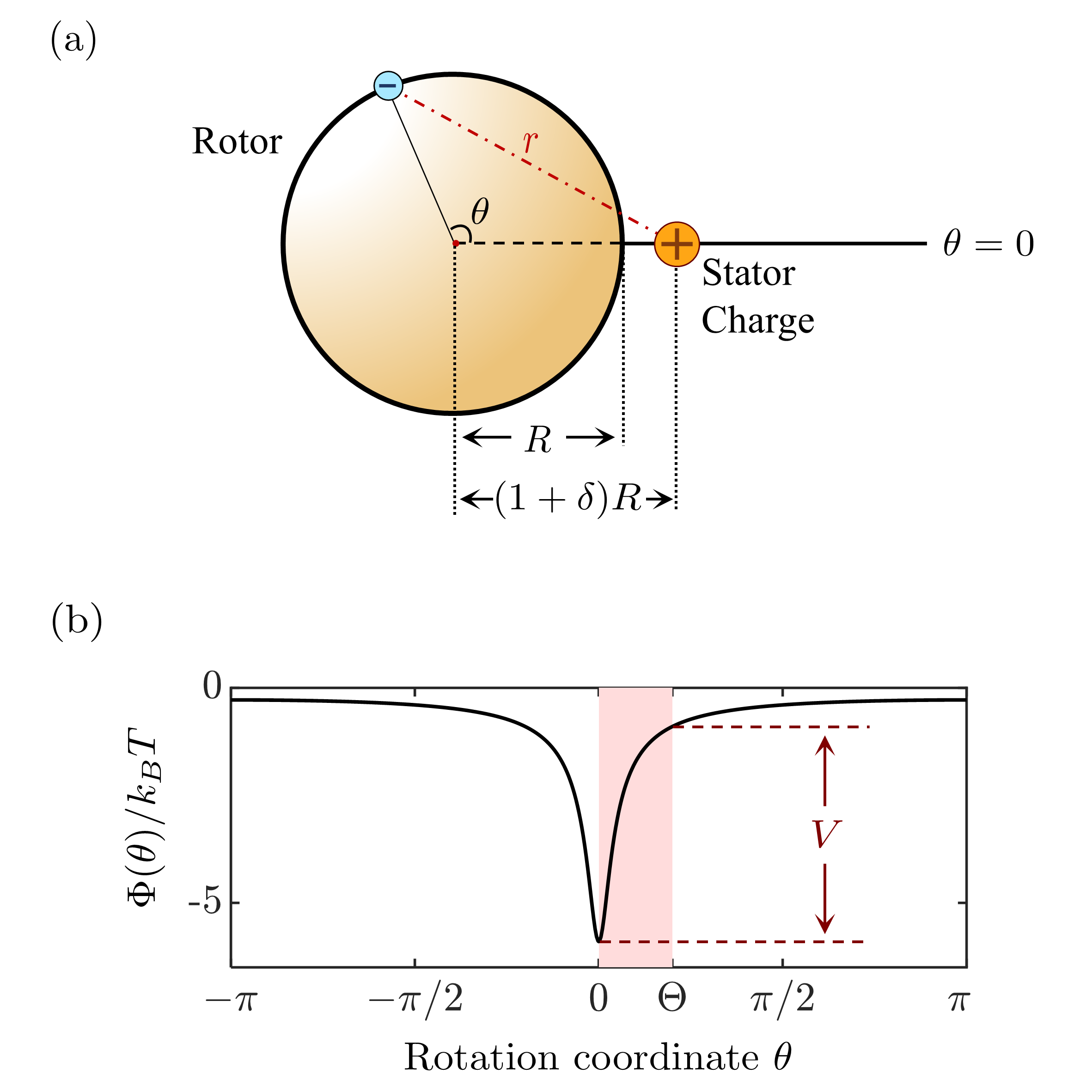}\caption{\protect\label{fig:Aux_Potential} Inside potential of the biological
motor. (a) A sketch of the structure of the microscopic rotor. The
radius of the rotor is denoted by $R$, and the distance between the
center of the rotor and the stator is $(1+\delta)R$. (b) The inside
potential $\Phi$ as a function of the angular position $\theta$.
The potential $\Phi(\theta)$ is induced by the Coulomb interaction
with a minimum at $\theta=0$. The red shaded region represents the
range where the site is in contact with the HCR. The inside potential
depth is defined as the drop of $\Phi$ over this region, i.e., $\Phi(\Theta)-\Phi(0)=V$.
In the plot, we have used parameters with the depth of the inside
potential as $V=5k_{B}T$ and $\Theta=\pi/5$.}
\end{figure}
\label{sec:Form-of-the} In this appendix, we derive the form of the
inside potential $\Phi(\theta)$ induced by the Coulomb interaction
between the stator charge and the binding site. In Fig. \ref{fig:Aux_Potential}(a),
we denote the radius of rotor by $R$, and the distance between the
stator charge and the center of the rotor by $(1+\delta)R$. The distance
between the binding site and the stator charge with respect to the
rotation coordinate $\theta$ is given by $r=R\sqrt{\delta^{2}+2(1-\cos\theta)\delta+2(1-\cos\theta)}.$
The potential energy of the Coulomb interaction between the stator
charge and the unbounded binding site is
\begin{equation}
\Phi(\theta)=-\frac{\mathcal{E}}{r}.\label{eq:aux_pot_form}
\end{equation}
where $\Phi_{0}$ defines the potential energy at infinity, and $\mathcal{E}\equiv q_{1}q_{2}/(4\pi\epsilon)$
with $q_{1},q_{2}$ as the charges for the binding site and stator
charge, $\epsilon$ the permittivity of the medium. In the main text,
we choose $\delta=0.1$, and the parameter $\mathcal{E}(V)$ is determined
by $\Phi(\Theta)-\Phi(0)=V$. The plot of the attractive inside potential
$\Phi(\theta)$ is illustrated in Fig. \ref{fig:Aux_Potential}(b)
for $V=5k_{B}T$. The red shaded region $(0\leq\theta<\Theta)$ represents
the range where the site is in contact with the high potential chemical
reservoir.

\section{$n-\Phi$ Diagram for the quasi static cycles}

\label{sec:n_Phi_diag}
\begin{figure*}
\centering{}\includegraphics{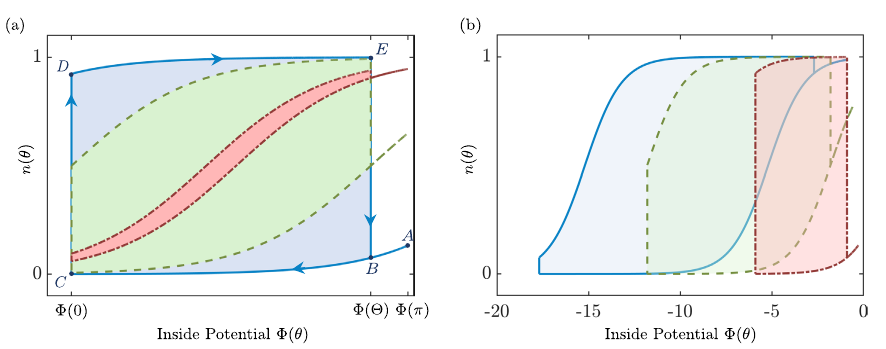}\caption{\protect\label{fig:quasi_static_diagram} $n-\Phi$ diagram for quasi
static cycles. (a) Cycles with different chemical potential differences:
$\Delta\mu=10k_{B}T$ (blue solid curve), $\Delta\mu=5k_{B}T$ (green
dashed curve) and $\Delta\mu=0.5k_{B}T$ (red dash-dotted curve).
The IPD is fixed as $V=5k_{B}T$ (b) Cycles with different IPDs: $V$
with values of $V=15k_{B}T$ (blue solid curve), $V=10k_{B}T$ (green
dashed curve), and $V=5k_{B}T$ (red dash-dotted curve). The chemical
potential difference is fixed as $\Delta\mu=10k_{B}T$.}
\end{figure*}

In this appendix, we provide a detailed examination
of the $n-\Phi$ diagram for quasi static cycles. In the $n-\Phi$
diagram, every point represents an equilibrium state of the binding
site. The $n-\Phi$ diagrams for quasi static cycles under different
values of $\Delta\mu$ are illustrated in Fig. \ref{fig:quasi_static_diagram}(a)
(the IPD is fixed as $V=5k_{B}T$). The blue solid curves, green dashed
curves, and red dash-dotted curves represent the $n-\Phi$ diagrams
for $\Delta\mu=10k_{B}T,5k_{B}T$ and $0.5k_{B}T$, respectively (the
chemical potential difference is fixed as $\Delta\mu=10k_{B}T$).
The $n-\Phi$ diagram for quasi static cycles under different values
of IPD $V$ are illustrated in Fig. \ref{fig:quasi_static_diagram}(b).
The blue solid curves, green dashed curves, and red dash-dotted curves
represent the $n-\Phi$ diagrams for $V=15k_{B}T,10k_{B}T$ and $5k_{B}T$,
respectively.

We show the $n-\Phi$ diagram for the case $\Delta\mu=10k_{B}T$ (blue
solid curve in Fig. \ref{fig:quasi_static_diagram}(a)) as an example
to describe the motor's cycling process. The thermodynamic cycle is
illustrated by the loop 
\[
A\rightarrow B\rightarrow C\rightarrow D\rightarrow E\rightarrow B\rightarrow A.
\]
The processes of the cycle are described as follows:
\begin{itemize}
\item $A\rightarrow B\rightarrow C$: Iso-chemical potential process with
$\mu=\mu_{L}$. In this process, the initial state $A$ of the binding
site corresponds to its position $\theta=-\pi$, and the final state
$C$ corresponds to $\theta=0$.
\item $C\rightarrow D$ : Isometric relaxation with $\Phi=\Phi(0)$. In
this process, the chemical potential of the binding site changes from
$\mu_{L}$ to $\mu_{H}$, combining with particles.
\item $D\rightarrow E$: Iso-chemical potential process with $\mu=\mu_{H}$.
During this process ($0\le\theta\le\Theta$), the binding site is
in contact with the HCR.
\item $E\rightarrow B$: Isometric relaxation with $\Phi=\Phi(\Theta)$.
In this process, the chemical potential of the binding site is changed
from $\mu_{H}$ to $\mu_{L}$, releasing particles.
\item $B\rightarrow A$: Iso-chemical potential process with $\mu=\mu_{L}$.
This process corresponds to the rotation from $\theta=\Theta$ to
$\theta=\pi$, restoring the binding site from the state $B$ to the
initial state $A$.
\end{itemize}
The irreversibility during the quasi static cycles is caused by the
relaxation processes $E\rightarrow B$ and $C\rightarrow D$, during
which the site exchanges particles with a reservoir having a different
chemical potential. When the chemical potential difference $\Delta\mu\rightarrow0$
{[}shown in Fig. \ref{fig:quasi_static_diagram}(a){]}, or the inside
potential depth $V\rightarrow\infty$ {[}shown in Fig. \ref{fig:quasi_static_diagram}(b){]},
the relaxation processes tend to vanish in the $n-\Phi$ diagram,
resulting in the cycle being approximately reversible.

\section{Power efficiency constraint relation in linear response regime}

\label{sec:Power-efficiency-constraint}In this appendix, we consider
the finite time behavior of the motor within the LR regime with $\Delta\mu\rightarrow0$.
In this situation, the rotation is slow enough to ensure a positive
work output, i.e., $\omega\tau_{r}\rightarrow0$, which will be validated
as self-consistent later. Thus, we can approximate $n(\theta)$ with
a series in terms of $\omega\tau_{r}$ as
\[
n(\theta)=\sum_{k=0}^{\infty}\left(-\omega\tau_{r}\frac{\mathrm{d}}{\mathrm{d}\theta}\right)^{k}n^{(0)}(\theta),
\]
which is a solution to Eq. (\ref{eq:evolution_eq}). By keeping $\omega\tau_{r}$
to the first order, the function is obtained as 
\[
n(\theta)=n^{(0)}(\theta,\mu_{H})-\frac{\omega\tau_{r}\beta\mathrm{e}^{-\beta(E_{b}-\Phi(\theta)-\mu_{H})}}{\left[1-\mathrm{e}^{-\beta(E_{b}-\Phi(\theta)-\mu_{H})}\right]^{2}}\Phi'(\theta),
\]
for $0<\theta\le\Theta$, and 
\[
n(\theta)=n^{(0)}(\theta,\mu_{L})-\frac{\omega\tau_{r}\beta\mathrm{e}^{-\beta(E_{b}-\Phi(\theta)-\mu_{L})}}{\left[1-\mathrm{e}^{-\beta(E_{b}-\Phi(\theta)-\mu_{L})}\right]^{2}}\Phi'(\theta),
\]
for $\Theta<\theta\le2\pi$.

Within such an LR regime, the net work output and particle number
exchange are expressed as series in $\omega$ and $\Delta\mu$, namely
\begin{align}
\Delta n\approx & \Delta n^{(0)}-\omega\tau_{r}\Delta n^{(1)},\label{eq:Delta_n_expand}\\
W_{\mathrm{net}}\approx & W^{(0)}-\omega(2\pi\gamma+\tau_{r}W^{(1)}),\label{eq:W_net_expand}
\end{align}
with
\begin{align*}
\Delta n^{(0)} & \approx\Omega^{(0)}+\Omega^{(1)}\beta\Delta\mu,\quad W^{(0)}=\Omega^{(0)}\Delta\mu,\\
\Delta n^{(1)} & \approx\left(\Omega^{(1)}+\Omega^{(2)}\beta\Delta\mu\right)\left[\Phi'(\Theta)-\Phi'(0)\right],\\
W^{(1)} & \approx\Lambda^{(0)}/\beta+\Lambda^{(1)}\Delta\mu.
\end{align*}
Here, $\Phi'$ represents the derivative of the inside potential $\Phi$
with respect to $\theta$, and the coefficients $\Omega^{(i)}(V)$
and $\Lambda^{(i)}(V)$ are functions of the inside potential depth
$V$. They are expressed as 
\begin{align*}
\Omega^{(0)} & =\frac{\sinh(\beta V/2)}{1+\cosh(\beta V/2)},\\
\Omega^{(1)} & =\frac{1}{4\cosh^{2}(\beta V/4)},\\
\Omega^{(2)} & =\frac{\mathrm{e}^{\beta V/2}(\mathrm{e}^{\beta V/2}-1)}{\left(\mathrm{e}^{\beta V/2}+1\right)^{3}},
\end{align*}
and
\[
\Lambda^{(0)}=\int_{0}^{2\pi}\mathrm{d}\theta\left[\frac{\beta\Phi'(\theta)}{2\cosh[\beta\left(\Phi(\theta)-\left\langle \Phi\right\rangle \right)^{2}/2]}\right]^{2},
\]
\[
\Lambda^{(1)}=\int_{0}^{\Theta}\mathrm{d}\theta\frac{\left[\beta\Phi'(\theta)\right]^{2}\mathrm{e}^{\beta\left(\Phi(\theta)-\left\langle \Phi\right\rangle \right)}\left(\mathrm{e}^{\beta\left(\Phi(\theta)-\left\langle \Phi\right\rangle \right)}-1\right)}{\left(\mathrm{e}^{\beta\left(\Phi(\theta)-\left\langle \Phi\right\rangle \right)}+1\right)^{3}}.
\]
The overall free energy change of the motor and the two reservoirs
is expressed as
\[
\Delta F=W_{\mathrm{net}}-T\Delta S^{(\mathrm{ir})},
\]
with Eqs. (\ref{eq:Delta_n_expand}) and (\ref{eq:W_net_expand}),
we obtain the irreversible entropy production as
\[
\Delta S^{(\mathrm{ir})}\approx\frac{2\pi(2\pi\gamma+\tau_{r}\Lambda^{(0)}/\beta)}{\tau_{\mathrm{cycle}}},
\]
which is inversely proportional to the cycle time $\tau_{\mathrm{cycle}}$,
indicating that our model represents a low dissipation chemical engine
\citep{esposito_efficiency_2010,Holubec2017}.

Now we consider (\ref{eq:W_net_expand}) with the positive work condition
of $W_{\mathrm{net}}>0$. There is 
\begin{align*}
\omega & <\frac{W^{(0)}}{2\pi\gamma+\tau_{r}W^{(1)}}=\frac{\Omega^{(0)}\Delta\mu}{2\pi\gamma+\tau_{r}\left(\Lambda^{(0)}/\beta+\Lambda^{(1)}\Delta\mu\right)}\\
 & \approx\frac{\Omega^{(0)}}{2\pi\gamma+\tau_{r}\Lambda^{(0)}/\beta}\Delta\mu.
\end{align*}
Thus, with given frictional coefficient $\gamma$, the positive work
condition ensures that $\omega\tau_{r}\Delta\mu$ is a higher-order
infinitesimal with respect to $\Delta\mu$, i.e., $\omega\tau_{r}\Delta\mu\sim O(\Delta\mu^{2})$.
Therefore, the assumption $\omega\tau_{r}\rightarrow0$ for $\Delta\mu\rightarrow0$
is consistently validated within our framework.

The efficiency $\eta$ is then obtained as 
\begin{align}
\eta & =\frac{W_{\mathrm{net}}}{\Delta\mu\Delta n}=\frac{\Omega^{(0)}\Delta\mu-\omega(2\pi\gamma+\tau_{r}\Lambda^{(0)}/\beta+\tau_{r}\Lambda^{(1)}\Delta\mu)}{\left(\Omega^{(0)}+\Omega^{(1)}\beta\Delta\mu\right)\Delta\mu},\nonumber \\
 & \approx\frac{\Omega^{(0)}\Delta\mu-\omega(2\pi\gamma+\tau_{r}\Lambda^{(0)}/\beta)}{\Omega^{(0)}\Delta\mu}.\label{eq:eta_omegaV}
\end{align}
With the approximation to the leading order, the power is obtained
as 
\begin{equation}
P\approx\frac{\omega}{2\pi}\left(\Omega^{(0)}\Delta\mu-\omega(2\pi\gamma+\tau_{r}\Lambda^{(0)}/\beta)\right).\label{eq:Power_omegaV}
\end{equation}
With Eqs. (\ref{eq:eta_omegaV}) and (\ref{eq:Power_omegaV}), $\omega$
is canceled, resulting in 
\[
P=\frac{\Delta\mu^{2}}{2\pi}\frac{\left[\Omega^{(0)}(V)\right]^{2}}{2\pi\gamma+\Lambda^{(0)}(V)\tau_{r}/\beta}\eta(1-\eta).
\]
Eqs. (\ref{eq:power_efficiency_constraint}) and (\ref{eq:MP}) are
derived by optimizing the IDP $V$ to achieve a maximum $P$ with
respect to the specific efficiency $\eta$.

Within such an LR regime, the rotation speed at maximum power is obtained
as
\[
\omega^{*}\approx\frac{\Omega^{(0)}\Delta\mu}{2\left(2\pi\gamma+\tau_{r}\Lambda^{(0)}/\beta\right)}.
\]
Hence, the average particle exchange rate between the two reservoirs
is linear with respect to the chemical potential difference, i.e.,
\[
J\equiv\omega\Delta n/2\pi\approx\frac{\left[\Omega^{(0)}\right]^{2}}{2\left(2\pi\gamma+\tau_{r}\Lambda^{(0)}/\beta\right)}\Delta\mu\propto\Delta\mu.
\]
The average effect of the steady rotation for the motor generates
a current whose response is linear with respect to the driving $\Delta\mu$
\citep{Proesmans2016,Raz2016,Holubec2017}.

\bibliographystyle{apsrev4-1}
\bibliography{Chemical_engine_refs}

\end{document}